# P*: A Model of Pilot-Abstractions

Andre Luckow[1], Mark Santcroos[2,1], Ole Weidner[1], Andre Merzky[1], Pradeep Mantha[1], Shantenu Jha[3,1*]

[1]*Center for Computation & Technology, Louisiana State University, USA*
[2]*Bioinformatics Laboratory, Academic Medical Center, University of Amsterdam, The Netherlands*
[3] *Rutgers University, Piscataway, NJ 08854, USA*
[*]*Contact Author:* `shantenu.jha@rutgers.edu`

*Abstract*—Pilot-Jobs support effective distributed resource utilization, and are arguably one of the most widely-used distributed computing abstractions – as measured by the number and types of applications that use them, as well as the number of production distributed cyberinfrastructures that support them. In spite of broad uptake, there does not exist a well-defined, unifying conceptual model of Pilot-Jobs which can be used to define, compare and contrast different implementations. Often Pilot-Job implementations are strongly coupled to the distributed cyberinfrastructure they were originally designed for. These factors present a barrier to extensibility and interoperability. This paper is an attempt to (i) provide a minimal but complete model (P*) of Pilot-Jobs, (ii) establish the generality of the P* Model by mapping various existing and well known Pilot-Job frameworks such as Condor and DIANE to P*, (iii) derive an interoperable and extensible API for the P* Model (Pilot-API), (iv) validate the implementation of the Pilot-API by concurrently using multiple distinct Pilot-Job frameworks on distinct production distributed cyberinfrastructures, and (v) apply the P* Model to Pilot-Data.

## I. Introduction and Overview

The seamless uptake of distributed infrastructures by scientific applications has been limited by the lack of pervasive and simple-to-use abstractions at multiple levels – at the development, deployment and execution stages [1]. Even where meaningful abstractions exist, the challenges of implementing them in an extensible, reliable and scalable manner, so as to support multiple applications are formidable. The lack of appropriate implementations has in fact resulted in "one-off" solutions that address challenges in a highly customized manner. Tools and implementations are often highly dependent on and tuned to a specific execution environment, further impacting portability, reusability and extensibility. Semantic and interface incompatibility are certainly barriers, but so is the lack of a common architecture and conceptual framework upon which to develop similar tools a barrier. This general state of affairs also captures the specific state of the abstractions provided by *Pilot-Jobs (PJ)*. There are a plethora of working definitions and roles for Pilot-Jobs; from our perspective, a Pilot-Job provides the ability to utilize a placeholder job as a container for a dynamically determined set of compute tasks.

Distributed cyber/e-infrastructure is by definition comprised of a set of resources that is fluctuating – growing, shrinking, changing in load and capability (in contrast to a static resource utilization model of traditional parallel and cluster computing systems). The ability to utilize a dynamic resource pool is thus an important attribute of any application that needs to utilize distributed cyberinfrastructure (DCI) efficiently. As a consequence of providing a simple approach for decoupling workload management and resource assignment/scheduling, PJ provide an effective abstraction for dynamic execution and resource utilization in a distributed context. Not surprisingly, Pilot-Jobs have been one of the most successful abstractions in distributed computing. The fundamental reason for the success of the Pilot-Job abstraction is that Pilot-Jobs liberate applications/users from the challenging requirement of mapping specific tasks onto explicit heterogeneous and dynamic resource pools. Pilot-Jobs thus shield applications from having to load-balance tasks across such resources. The Pilot-Job abstraction is also a promising route to address specific requirements of distributed scientific applications, such as coupled-execution and application-level scheduling [2], [3].

A variety of PJ frameworks have emerged: Condor-G/Glide-in [4], Swift [5], DIANE [6], DIRAC [7], PanDA [8], ToPoS [9], Nimrod/G [10], Falkon [11] and MyCluster [12], to name a few. Although they are all, for the most parts, functionally equivalent – they support the decoupling of workload submission from resource assignment – it is often impossible to use them interoperably, or even just to compare them functionally or qualitatively.

Our objective is to provide a minimal, but complete model for Pilot abstractions [13] – referred to as P* Model ("P-star"), which we present in §II. The P* Model provides a conceptual basis to compare and contrast different PJ frameworks – which to the best of our knowledge is the first such attempt. We also investigate generalizations to the base P* Model: a natural and logical extension of the P* Model arises from the opportunity to extend it to include data in addition to computational tasks. This leads to an abstraction analogous to the Pilot-Job: the *Pilot-Data (PD)* abstraction. The potentially consistent treatment of data and compute suggests symmetrical compute and data elements in the model; thus we refer to this model as the P* Model of Pilot Abstractions.

In §III we validate the P* Model by analyzing well-known PJ frameworks (BigJob, Condor-G/Glide-in, DIANE) and mapping them to the elements of the P* Model. §IV of this paper motivates and describes the Pilot-API; we discuss how existing and widely used Pilot-Job frameworks can be used through the Pilot-API. §V describes the experiments and performance measurements used to characterize the workings of the Pilot-API and to demonstrate interoperability across middleware and infrastructure. To further substantiate the impact of P*, we will demonstrate interoperability between different PJ frameworks. We believe this is also the first demonstration of concurrent interoperation of different Pilot-Job imple-mentations. Performance advantages arising from the ability to distribute part of a data-intensive workload are discussed; interoperable capabilities increase flexibility in resource selection and optimization.

It is worth noting that Pilot-Jobs are used on every major national and international DCI, including NSF/XSEDE, NSF/DOE Open Science Grid (OSG), European Grid Initiative (EGI) and others; they are used to support hundreds of thousands of tasks daily. Thus we believe the impact and validation of this paper lies in its ability to not only influence but also bridge the theory and practice of Pilot-Jobs, and thus multiple domains of science dependent on distributed cyberinfrastructure.

## II. THE P* MODEL OF PILOT-ABSTRACTIONS

An initial motivation for the P* Model of pilot-abstractions is to provide a common analytical framework to understand commonly used Pilot-Job frameworks. The P* model was derived by analyzing many Pilot-Job implementations. We first present the common *elements* of the P* Model, followed by a description of the *characteristics* that determine the interaction of these elements and the overall functioning of any Pilot-Job framework consistent with the P* Model.

Before we proceed to discuss the P* Model, it is important to emphasize that there exists a plethora of terms (abstraction, model, framework, implementation etc) that are overloaded and overlapping, and often used inconsistently in the literature; thus we establish context and usage of relevant terms.

*Terms and Usage:* A Pilot-Job can be defined as an *abstraction* that generalizes the reoccurring concept of utilizing a placeholder job as a container for a set of compute tasks; an instance of that placeholder job is commonly referred to as *Pilot-Job* or *pilot*. The P* *model* provides a comprehensive description of the Pilot-Job abstraction, based on a set of identified elements and their interactions. The P* Model can be used as a *conceptual model*, for analyzing different implementations of the Pilot-Job abstraction. The *Pilot-API* provides an interface that exposes a sub-set of the P* elements and characteristics to applications. It is important to distinguish P* – which provides a conceptual (abstract) model, from an implementation of the P* Model. A *Pilot-Job framework* refers to a specific instance of a Pilot-Job implementation that provides the complete Pilot-Job functionality (e. g. BigJob).

### A. Elements of the P* Model

This sub-section defines the elements of the P* Model:
• **Pilot (Pilot-Compute):** The Pilot is the entity that actually gets submitted and scheduled on a resource. The Pilot provides application (user) level control and management of the set of allocated resources.
• **Compute Unit (CU):** A CU encapsulates a self-contained piece of work (a compute task) specified by the application, that is submitted to the Pilot-Job framework. There is no intrinsic notion of resource associated with a CU.

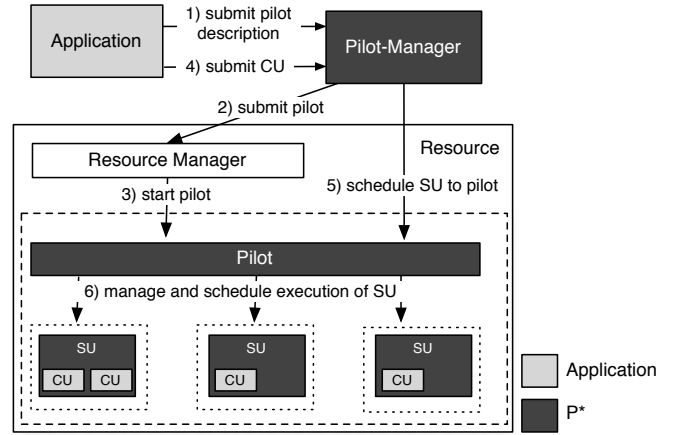

*Fig. 1:* **P\* Model: Elements, Characteristics and Interactions:** *The manager has two functions: it manages 1) Pilots (step 1-3) and 2) the execution of CUs. After a CU is submitted to the manager, it transitions to an SU, which is scheduled to a Pilot by the PM.*

• **Scheduling Unit (SU):** SUs are the units of scheduling, internal to the P* Model, i.e., it is not known by or visible to an application. Once a CU is under the control of the Pilot-Job framework, it is assigned to an SU.
• **Pilot-Manager (PM):** The PM is responsible for (i) orchestrating the interaction between the Pilots as well as the different components of the P* Model (CUs, SUs) and (ii) decisions related to internal resource assignment (once resources have been acquired by the Pilot-Job). For example, an SU can consist of one or more CUs. Further, CUs and SUs can be combined and aggregated; the PM determines how to group them, when SUs are scheduled and executed on a resource via the Pilot, as well as how many resources to assign to an SU.

The application utilizes a PJ framework to execute multiple instances (ensemble) of an application kernel (kernel: the actual binary that gets executed), or alternatively instances of multiple different application kernels (a workflow). To execute an application kernel, an application must define a CU specifying the application kernel as well as other parameters. This CU is then submitted to the PM (the entry point to the Pilot-Job framework), where it transitions to an SU. The PM is then responsible for scheduling the SU onto a Pilot and then onto a physical resource. As we will see in §III, the above elements can be mapped to specific entities in many existing Pilot-Job frameworks – more than one logical element are often rolled into one specific entity in a Pilot-Job.

### B. Characteristics of P* Model

We propose a set of fundamental properties/characteristics that describe the interactions between the elements, and thus aid in the description of P* Model.

**Coordination Characteristics:** these describe how various elements of the P* Model, i.e. the PM, the Pilot, the CUs and the SUs, interact. A common coordination pattern is master/worker (M/W): the PM represents the master process that controls a set of worker processes, the Pilots. The point of decision making is the master process. In addition to the *centralized* M/W, M/W can also be deployed *hierarchically*. Al-ternatively, coordination between the elements, in particular the Pilots, can be performed so as to be *decentralized*, i.e. without central decision making point.

**Communication Characteristics:** The communication characteristics describes the mechanisms for data exchange between the elements of the P* Model: e.g. messages (point-to-point, all-to-all, one-to-all, all-to-one, or group-to-group), streams (potentially unicast or multicast), publish/subscribe messaging or shared data spaces.

**Scheduling Characteristics:** describe the process of mapping a SU to resources via a Pilot, and potential multiple levels of scheduling. Scheduling has a spatial component (which SU is executed on which Pilot?) but also a temporal component (when to bind?). For example, for the temporal component: a SU can be bound to a Pilot either before the Pilot has in turn been scheduled (*early* binding), or binding occurs if the SU is bound after the Pilot has been scheduled (*late* binding). The different scheduling decisions that need to be made are representative of multi-level scheduling that is often required in distributed environments. For example, the Pilot is scheduled using the *system-level* scheduler. Once resources are assigned to the Pilot, *application-level* scheduling can occur at several levels both inside and outside the PJ framework.

The term *agent*, although not a part of the P* Model, finds mention when discussing implementations. For the purposes of this paper, an agent refers to a "proxy process" that has some decision making capability, and could aid the implementation of one or more of the characteristics of the P* Model (coordination, communication, scheduling), within a Pilot-Job framework. These agents can be used to enforce a set of (user-defined) policies (e.g. resource capabilities, data-/compute affinities, etc.) and heuristics.

### C. P* as a Model for Pilot-Data

Many scientific applications have immense data requirements, which are projected to increase dramatically in the near future [14]. While Pilot-Jobs efficiently support late-binding of *Compute Units* and resources, the analogous management of data in distributed systems remains a challenge due to various reasons: (i) the placement of data is often decoupled from the placement of Compute Units and Pilots, i.e. the application must often manually stage in and out its data using simple scripts; (ii) heterogeneity, e.g. with respect to storage, filesystem types and paths, often prohibits or at least complicates late binding decisions; (iii) higher-level abstraction that allow applications to specify their data dependencies on an abstract, logical level (rather than on file basis) are not available; (iv) due to lack of a common treatment for compute and data, optimizations of data/compute placements are often not possible.

This motivates an analogous abstraction that we call *Pilot-Data (PD)*. PD provides late-binding capabilities for data by separating the allocation of physical storage and application-level data units. Further, it provides an abstraction for expressing and managing relationships between data units and/or compute units. These relationships are referred to as *affinities*.

*P* Model Elements for Data:* The elements defined by P* can be extended with the following elements:
- **Pilot (Pilot-Data):** A Pilot-Data (PD) functions as a placeholder object that reserves the space for data units. PD facilitates the late-binding of data and resource and is equivalent to the Pilot in the compute model.
- **Data Unit (DU):** DU is the base unit of data assigned by the application, e.g. a set of data files.
- **Scheduling Unit (SU):** is an internal unit of scheduling (as in the compute case). The Pilot framework can aggregate or split DUs into one or more SUs.
- The **Pilot-Manager (PM)** is the same as in the compute model and implements the different characteristics of the P* Model. It is responsible for managing DUs and SUs. Data is submitted to the framework via the PM. The PM which is responsible for mapping DUs to SUs and for conducting decision regarding resource assignments. SUs are placed on physical resources via the Pilot.

Note, each element can be mapped to an element in the P* Model by symmetry, e.g., a DU correspond to a CU in the original P* Model; a PD is a placeholder reserving a certain amount of storage on a physical resource and corresponds to the Pilot in the P* Model.

*P* Model Characteristics for Data:* While the extended P* Model introduces new elements, the characteristics however, remain the same to a great extent. The coordination characteristic describes how the elements of PD interact, e.g. utilizing the M/W model; the communication characteristic can be applied similarly. The scheduling characteristics must be extended to not only meet compute requirements, but also to support common data patterns. The scheduling component particularly needs to consider affinities, i.e. user-defined relationships between CUs and/or DUs. Data-data affinities e.g. exist if different DUs must be present at the same resource; data-compute affinities arise if data and compute must be co-located – if their current location is different, data and compute placement decisions are made by the scheduler based on defined policies, affinities and dynamic resource information.

### D. Putting it all together

Figure 1 illustrates the interactions between the elements of the P* Model. The figure focuses on Pilot-Compute, for simplicity, but immediately applies to Pilot-Data by symmetry. First, the application specifies the capabilities of the required resources using a Pilot-Job description (step 1). The PM then submits the necessary number of Pilots to fulfill the resource requirements of the application (step 2). Each Pilot is queued at a resource manager, which is responsible for starting the Pilot (step 3). There can be variations of this flow: while in the described model, the application defines the required resources, the PM could also decide based on the submitted CU workload whether and when it submits new Pilots.

The application can submit CUs to the PM at any time (step 4). A submitted CU becomes an SU, i.e. the PM is now in control of its scheduling. In the simplest case one CU corresponds to one SU; however, SUs can be combined and aggre-gated to optimize throughputs and response times. Commonly, a hierarchical M/W model for coordination is internally used: the PM uses M/W to coordinate a set of Pilots, the Pilot itself acts as manager for the execution of the assigned SUs.

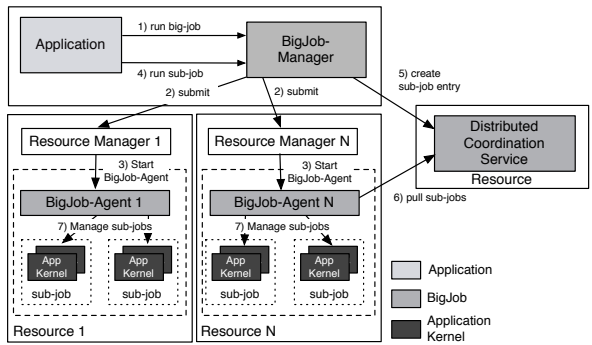

*Fig. 2: **BigJob Architecture and Mapping to P\*:*** The BJ architecture resembles many elements of the P* Model. The BigJob-Manager is the central Pilot-Manager, which orchestrates a set of Pilots. Each Pilot is represented by a decentral component referred to as the BigJob-Agent. Sub-job – the CUs– are submitted via the PM. CUs are mapped 1:1 to SUs.

Scheduling decisions can be made on multiple levels. A Pilot is bound to a physical resource on which it is responsible for a particular resource set. The PM is responsible for selecting a Pilot for an SU (step 5). Once a SU has been scheduled to a Pilot, the Pilot decides when and on which part of the resource an SU is executed. Further, the Pilot manages the subsequent execution of an SU (step 6). There can be variations of this flow. PJ frameworks with decentralized decision making e. g. often utilize autonomic agents that pull SUs according to a set of defined policies.

## III. PILOT-JOB FRAMEWORKS

The aim of this section is to provide a basic understanding of some of the most commonly used PJ frameworks. This will serve to both motivate the development of the P* Model as well as validate it. In particular, we focus on Condor-G/Glide-in, BigJob and DIANE, and show that the P* Model can be used to explain/understand these and other PJ frameworks.

### A. Condor-G/Glide-in

The Condor project pioneered the concept of Pilot-Jobs by introducing the *Condor-G/Glide-in* mechanisms [4] which allow the temporary addition of Globus GRAM controlled HPC resources to a Condor resource pool. The Pilot is exposed as a complete Condor pool that is started via the Globus GRAM service of a resource. This mechanism is referred to as Condor Glide-in. Subsequently, jobs (CUs) can be submitted to the Condor Glide-in pool via standard Condor tools and APIs.

GlideinWMS [15] is a higher-level workload management system built on top of the Pilot capabilities of Condor-G/Glide-in. The system can based on the current and expected number of jobs in the pool, automatically increase or decrease the number of active Glide-ins (Pilots) available to the pool. In contrast to Condor-G/Glide-In, the Pilot capabilities are not directly exposed to the end-user. GlideinWMS is the recommended mode for accessing OSG resources.

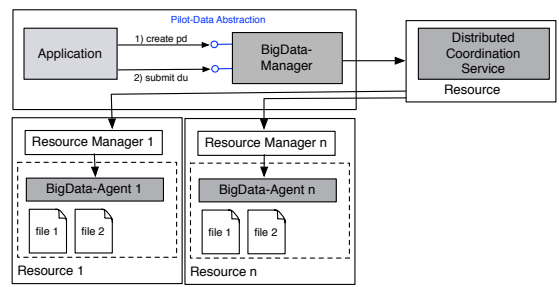

*Fig. 3: **BigData Architecture and Interactions:*** Similar to BJ, a BD-Manager orchestrates a set of Pilots, i. e. BigData-Agent. Coordination is carried out via a distributed coordination service.

### B. DIANE

DIANE [6] is a task coordination framework, which was originally designed for implementing master/worker applications, but also provides PJ functionality for job-style executions. it utilizes a single hierarchy of worker agents and a PJ manager referred to as `RunMaster`. For the spawning of PJs a separate script, the so-called submitter script, is required. For access to the physical resources the GANGA framework [16] can be used. Once the worker agents are started they register themselves at the RunMaster. For communication between the RunMaster and worker agents point-to-point messaging based on CORBA is used. CORBA is also used for file staging. DIANE includes a simple capability matcher and FIFO-based task scheduler. Plugins for other workloads, e. g. DAGs or for data-intensive application, exist or are under development.

### C. BigJob and BigData: A SAGA-based PJ Framework

BigJob (BJ) [17], [18] is a SAGA-based PJ framework. BJ has been designed to be general-purpose and extensible. While BJ has been originally built for HPC infrastructures, such as XSEDE and FutureGrid, it is generally also usable in other environments. This extensibility mainly arises from the usage of SAGA [19], [20] as a common API for accessing distributed resources. Figure 2 illustrates the BJ architecture and its mapping to P*. The architecture reflects all P* elements: The BJ-Manager is the Pilot-Manager responsible for coordinating the different components of the frameworks. The BigJob-Agent is the actual Pilot that is submitted to a resource. CUs are referred to as sub-jobs. Internally CUs are mapped 1:1 to SUs.

BJ implements the following P* characteristic: As coordination model the M/W scheme is used: The BJ-Manager is the central entity, which manages the actual Pilot, the BJ-Agent. Each agent is responsible for gathering local information, for pulling sub-jobs from the manager, and for executing SUs on its local resource. The SAGA Advert Service is used for communication between manager and agent. The Advert Service (AS) exposes a shared data space that can be accessed by manager and agent, which use the AS to realize a push/pull communication pattern, i. e. the manager pushes a SU to the AS while the agents periodically pull for new SUs. Results and state updates are similarly pushed back from the agent to the manager. Further, BJ provides a pluggable communication & coordination layer and also supports other c&c systems besides the AS, e. g. Redis [21] and ZeroMQ [22].

| P* Element | BigJob | DIANE | Condor-G/ Glide-in |
|---|---|---|---|
| Pilot-Manager | BigJob Manager | RunMaster | condor_master condor_collector condor_negotiator condor_schedd |
| Pilot | BigJob Agent | Worker Agent | condor_master condor_startd |
| Compute Unit (CU) | Task | Task | Job |
| Scheduling Unit (SU) | Sub-Job | Task | Job |

TABLE I: **Mapping P* elements and PJ Frameworks:** *While each PJ framework maintains its own vocabulary, each of the P* elements can be mapped to one (or more) components of the different PJ frameworks.*

*BigData (BD)* is an implementation of the Pilot-Data abstraction. BigData is designed as an extension of BigJob. Figure 3 provides an overview of the architecture of BigData. Similar to BigJob, it is comprised of two components: the BD-Manager and the BD-Agents, which are deployed on the physical resources. The coordination scheme used is Master-Worker (MW), with some decentralized intelligence located at the BD-Agent. The BD-Manager is responsible for (i) metadata management, i.e. it keeps track of all PD and associated DUs, (ii) for scheduling of data movements and replications (taking into account the application requirements defined via affinities), and (iii) for managing data movements activities. BigData supports plug-able storage adaptors (currently for SSH, WebHDFS [23] and Globus Online [24]).

### D. Discussion

P* provides an abstract model for describing and understanding PJ frameworks, i.e., different components of PJ frameworks can be mapped to the P* elements. While each of the frameworks maintains its own vocabulary, all share the common P* elements. Table I summarizes how P* can be applied to BigJob, DIANE and Condor-G/Glide-in. Table II summarizes the P* characteristic and other properties of these frameworks. While most of these frameworks share many properties, such as the M/W coordination model, they differ in characteristics, such as communication model or scheduling. Despite of the many commonalities, the different PJ frameworks have different usage modalities mainly cause by the fact that each PJ framework has evolved in a specific infrastructures, e.g., Condor-G/Glide-in is the native PJ framework of the Open Science Grid.

## IV. PILOT-API: A UNIFORM API TO HETEROGENEOUS PJ FRAMEWORKS

In the previous two sections we presented successively the P* Model and existing Pilot-Job frameworks. Before we present the Pilot-API – which provides an abstract interface to Pilot-Job frameworks that adhere to the P* Model, we will motivate the need for such an API.

### A. Motivation

At a high-level, there exist two approaches towards interoperability: (i) deep integration of systems (system level interoperability), and (ii) the use abstract interfaces (application level interoperability). Approach (i) requires a certain level of semantic harmonization between the systems, and is (in principle and technically) hard to achieve post-facto, even if the respective systems inherently implement the same abstract model (here: the P* Model). While interoperation via an abstract interface (ii) (here: Pilot-API) is a semantically weaker approach than (i), it does allow for interoperability with minimal (application level) effort [18], [25].

| Properties | BigJob | DIANE | Condor-G/ Glide-in |
|---|---|---|---|
| **Coordination** | M/W | M/W | M/W |
| **Communication** | Advert Service | CORBA | TCP |
| **Scheduling** | FIFO, custom | FIFO, custom | Matchmaking, priority-based scheduler |
| Agent Submission | API | GANGA Submission Script | Condor CLI |
| End User Environment | API | API and M/W Framework | CLI Tools |
| Fault Tolerance | Error propagation | Error propagation, retries | Error propagation, retries |
| Resource Abstraction | SAGA | GANGA/ SAGA | Globus |
| Security | Multiple (GSI, User/Pass.) | Multiple (GSI) | Multiple (GSI, Kerberos) |

TABLE II: **P* Characteristics and Properties of Different Pilot-Job Frameworks:** *The properties in bold-face correspond to the P* characteristics; other items are general properties. The PJ frameworks share many P* characteristics and properties, e.g. the common usage of the M/W scheme or of a resource abstraction layer. However, they also differ in aspects, such as the coordination model or the communication framework.*

We appreciate the difficulty of designing an API for multiple, heterogeneous systems with the right level of semantic expressivity and simplicity [26]. Defining the API as 'smallest common denominator' is often too simplifying and misses large numbers of 'edge' use cases; defining the API as 'greatest common factor' clutters the API with non-portable semantics, making the API too complex [27]. The Pilot-API uses the Pareto principle as guideline for a balanced abstraction level.

### B. Understanding the Pilot-API

The Pilot-API [28] is a Python-based API and supports two different usage modes (i) it provides a unified API to various PJ frameworks (e.g. BigJob, DIANE and Condor-G/Glide-in), and (ii) it enables the concurrent usage of multiple PJ frameworks. The Pilot-API classes and interactions are designed to reflect the P* elements and characteristics. The API exposes the two primary interfaces: the `PilotComputeService` is responsible for the management of Pilots and the `ComputeUnitService` for the management of CUs. As defined by P*, a CU represents a primary self-containing piece of work that is submitted through the Pilot-API.

Figure 4 shows the interactions between the Pilot-API entities. The Pilot-API decouples workload management and resource scheduling by exposing two separate services: The `PilotComputeService` and `ComputeUnitService`. The `PilotComputeService` serves as a factory for creating Pilots. Also, it can be used to query for currently active `PilotCompute` instances. A `PilotCompute` instance is returned as result of the `create_pilot()` method of the `PilotComputeService` (step 1). The instantiation of the `PilotCompute` instance is done by using a `PilotComputeDescription`. The description can be reused and has no state, while the `PilotCompute` instance has state and is a reference for further usage: the references `PilotCompute` object represents a Pilot instance and allows the application to interact with it, e.g. to query its state or to cancel it. The process of `PilotCompute` creation is depicted in step 1-2 of Figure 4 and in Listing 1.

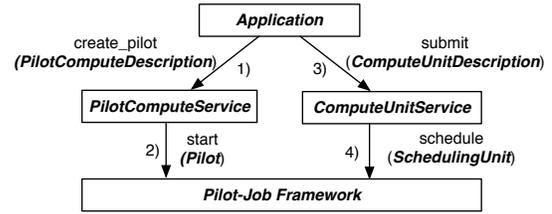

Fig. 4: **Control Flow Pilot-API and PJ Frameworks:** The functionality of pilot-jobs are exposed using two primary classes: The `PilotComputeService` for the management of Pilots, and the `ComputeUnitService` for the management of CUs.

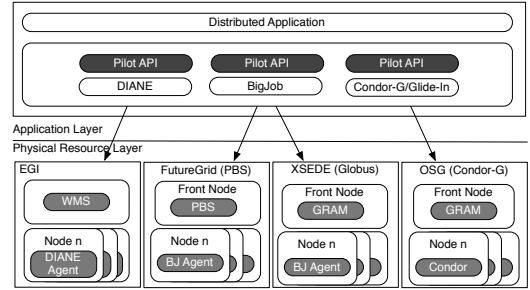

Fig. 5: **Pilot-API and PJ frameworks:** The Pilot-API provides a unified interface to utilize the native Pilot-Job capabilities of different infrastructures, e.g. BigJob for XSEDE/FutureGrid, DIANE for EGI and Condor for OSG.

```
pcs = PilotComputeService()
pc_desc = PilotComputeDescription()
pc_desc.total_core_count = 8
pc = pcs.create_pilot('gram://queenbee',
                      pc_desc, 'bigjob')
```

Listing 1: *Creation of a `PilotCompute` instance using a `PilotComputeDescription`.*

Listing 2 shows the creation of a `ComputeUnitService`. Having created a `ComputeUnitService` instance, `PilotComputeService` instances can be added and removed at any time, which makes the respective pilots available to that Service. These semantics enable applications to respond to dynamic resource requirements at runtime, i.e. additional resources can be requested on peak demands, and can be released if they are no longer required.

```
cus = ComputeUnitService()
cus.add(pcs)
```

Listing 2: *Instantiation of a `ComputeUnitService` using a reference to the `PilotComputeService`.*

The `ComputeUnitService` is responsible for managing the execution of CUs. Regardless of the state of the `PilotComputeService`, applications can submit CUs to a `ComputeUnitService` at anytime (Listing 3 and step 3 in Figure 4). Once the `ComputeUnitService` becomes responsible for a CU, the CU transitions to an SU. SUs are internally processed (e.g. they can be aggregated) and are then scheduled to the Pilot-Job Framework (step 4). The PJ framework is responsible for the actual execution of the SU on a resource. Note that multiple levels of (hierarchical) scheduling can be present – commonly a SU is scheduled inside a PJ framework, the model allows it to be present in multiple layers.

```
cud = ComputeUnitDescription()
cud.executable = '/bin/bfast'
cud.arguments = ['match', '-t4', '/data/file1']
cud.total_core_count = 4
cu = cus.submit(cud)
```

Listing 3: *Instantiation and submission of a `ComputeUnitDescription`.*

Each `ComputeUnit` and `PilotCompute` object is associated with a state. The state model is well-defined. Applications can query the state using the `get_state()` method or they can subscribe to state update notifications using callbacks.

Finally, an application can have any number of `PilotComputeService` or `ComputeUnitService` instances. Multiple `PilotComputeService` instances can be associated to a `ComputeUnitService`, and a `PilotComputeService` can be associated to multiple `ComputeUnitService` instances. A `ComputeUnitService` can manage multiple `ComputeUnit` instances, but a `ComputeUnit` can only be managed by one `ComputeUnitService`. Similarly, a `PilotComputeService` can manage multiple `PilotCompute` instances, but a `PilotCompute` can only be managed by one `PilotComputeService`.

### C. Pilot-API for Data

Analogous to the Pilot-API for Compute, the Pilot Data API [28] defines the `PilotDataService` entity as an abstraction for creating and managing pools of storage. A `PilotData` instance represents the actual physical storage space. An additional `ComputeDataService` entity functions as an application-level scheduler, which accepts both `ComputeUnits` and `DataUnits` – this resolves new dependencies (e.g. data/data or data/computer affinities), and is responsible for managing the execution of DUs and CUs.

In summary, the Pilot-API provides a well-defined abstraction for managing both compute and data. The API has been developed to support production-scale science on production infrastructure. As shown in Figure 5 the API supports different HPC and HTC infrastructures.

## V. EXPERIMENTS AND RESULTS

As discussed in §III, several PJ frameworks can be collectively used via the Pilot-API. We begin by understanding the overhead of PJ frameworks (section V-A). In section V-B we show the effectiveness of the Pilot-API/P* Model approach by executing *real application workloads* – a genome alignment application – on multiple distinct production (XSEDE, EGI, OSG) and research (FutureGrid) infrastructures. It is important to note that our experiments do not try to identify the "fastest" PJ framework, as this is dependent on several external factors, often specific to the infrastructure used. In stead, we focus on demonstrating interoperability via the common Pilot-API by using multiple PJ frameworks concurrently on multiple infrastructures (see section V-C). Further, we show how Pilot-Data enables applications (i) to lower the volume of the transferred data, and (ii) to utilize different transfer protocols in section V-D.

### A. PJ Frameworks Overhead

Before we understand the performance of different frameworks for real application workloads, we analyze the typical overheads for BigJob and DIANE. The overhead of a PJ framework, like many distributed submission mechanisms and tools, is most commonly determined by the following factors: the API overhead, the job submission and coordination overhead. Although important determinants of the time-to-solution, the queueing and file staging time heavily depend on extrinsic factors, such as the system and network load, but are not intrinsic overheads of the PJ framework. The overhead of the API-layer, i. e. the Pilot-API, was not measurable using the built-in Python profiler, which provides accuracies in the magnitude of milliseconds; this places an effective upper-bound on Pilot-API latencies. Ref. [29] established that even in a distributed context, job submission overheads, are very low when compared to the runtimes in consideration. There are many factors that influence the overall performance, e. g. the degree of distribution (local (LAN) vs. remote (WAN)), the communication pattern (1:n versus n:n) and the communication frequency. We focus our investigation on the the communication & coordination (c&c) subsystem, which we established earlier as import characteristics of PJ frameworks.

We executed a different number of very short running (i. e. zero workload) CUs on Alamo/FG concurrently. This enables us to focus on the overhead induced by the c&c subsystem. In general, the c&c systems used are mostly insensitive to the number of coordinated CUs. Although we do not provide a detailed discussion of the dependency between coordination overhead and the number of CUs, it is worth mentioning that the runtime increases only slightly with the size of the pilot and/or the number of managed CUs.

Figure 6 illustrates the scalability of BJ and DIANE with respect to the number of cores and CUs managed by Pilot. For this purpose, we execute 4 CUs per core, i. e. between 32 and 512 CUs. BigJob with Redis (local) shows almost linear scaling up to 128 cores. BigJob with Redis (remote) imposes an increase of about 14 %. BigJob with ZeroMQ performs very well with lower core counts; with larger core counts, the runtimes increase, indicating a potential scalability bottleneck. Due to higher startup overhead, at lower core counts DIANE shows a longer runtime than ZeroMQ or Redis. At higher core counts DIANE behaves similar to BigJob/ZeroMQ, but shows a greater increase in the overall runtime. This increase is likely attributable to the single central manager in DIANE's CORBA-based client-server architecture. Using Redis as central data space for BigJob decouples Pilot-Manager and Agent, yielding better performance in particular with many CUs.

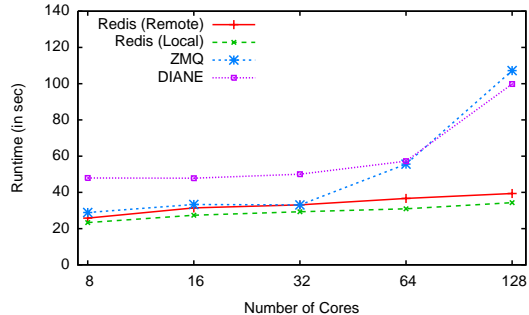

*Fig. 6:* **Pilot-Job Coordination Mechanism:** *The runtime of a workload of 4 CUs per core, i. e. 32 - 512 CUs, using different Pilots and configuration. For BJ-Redis the runtime increases only moderately, the client-server-based implementations BJ-ZMQ and CORBA-based DIANE show particularly a steep increase when going from 64 to 128 cores.*

### B. Characterizing PJ Frameworks on DCI

To validate the effectiveness and usability of the Pilot-API, we conducted a series of experiments on various production infrastructures. We executed BFAST [30] using three different PJ frameworks (BigJob, DIANE and Condor) on XSEDE [31], FutureGrid [32], EGI [33] and OSG [34]. Specifically, we utilized the following resources: XSEDE: Trestles (100 TFlop / 324 nodes / 10,368 cores / Torque) and QueenBee (Linux Cluster / 668 nodes / 5,344 cores / PBS); FutureGrid: India and Sierra (108 nodes / 864 cores / PBS); EGI: Resource federation of 364,500 cores; OSG: Condor pool (via the *Engage* VO, GlideinWMS, 20,000 Glide-ins).

*Experimental Configuration:* We run experiments using five different configurations: (B1) BigJob/XSEDE, (B2) BigJob/-FutureGrid, (B3) two BigJob/Futuregrid, (B4) DIANE/EGI, (B5) Condor/OSG. As discussed, the Pilot-API provides a unified interface for accessing these infrastructures using the respective native PJ framework, i. e. BigJob, DIANE and Condor (see figure 5). For BJ we use the PBS/SSH plugin to access both the FutureGrid and XSEDE machines. On OSG, we use SAGA and the SAGA-Condor adaptor to interface directly with OSG's dynamic GlideinWMS resource pool. Further, we utilize DIANE on EGI.

The investigated workload consists of 128 CUs. Each CU executes a BFAST matching process, which is used to find potential DNA sequence alignments. Each CU requires the following input data: 1.9 GB for the reference genome and index files, and 170 MB for the *short read* file (generated by the DNA sequencing machine). A total of 128 read files (one per CU) are used. Each BFAST CU requires 1 core; All input files are staged previous to the actual run.

*Experimental Results:* Figure 7 shows the results of the experiments. We measured the time to transfer input files ($T_X$), and the compute time $T_C$. $T_C$ also includes the overhead of the pilot. The total runtime ($T_R$) is the sum of $T_X$ and $T_C$.

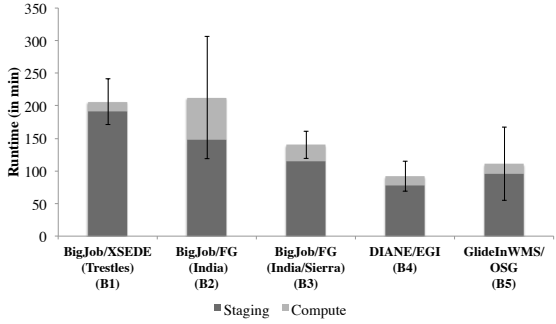

*Fig. 7: **PJ Framework Performance on XSEDE, FutureGrid, EGI and OSG:** Average runtime of 128 BFAST match tasks on 128 cores. Each experiment is repeated at least 3 times.*

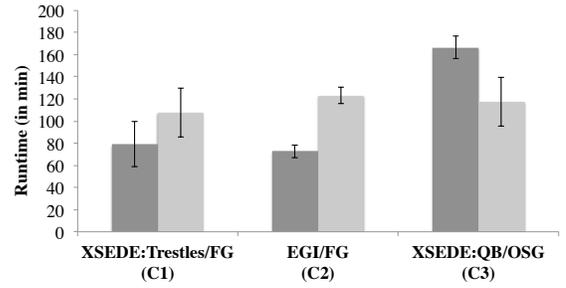

*Fig. 8: **PJ Framework Interoperability:** Average runtime of 128 BFAST CUs on different infrastructures. CUs are equally distributed across the two infrastructures. The performance heavily depends on the available bandwidths to the resources, which determines the time required for file transfers.*

For each CU, the reference genome, the index files and one read file need to be staged (1.9 GB); this constitutes the most significant part of the overall runtime. In particular, on HPC clusters, the staging quickly becomes a bottleneck due to the fact that all CUs share the incoming network bandwidth. Also, fluctuation of the bandwidth in this scenario leads to a high failure rate during downloads. As seen in B3 (middle bar), the distribution of the network load to two resources leads to a reduction of $T_X$. Also, in the HTC configuration (B4, B5), CUs are distributed across multiple machines minimizing the possibility of network congestions of single links. As an indication, a typical run of 128 CUs on OSG ended up being scheduled on 10+ sites spread over an equal amount of nodes per site.

In general $T_C$, is heavily dependent on the available disk I/O. Both Trestles (XSEDE) and India/Sierra (FG) have shared network filesystems (Lustre for Trestles, NFS for India), which are utilized by all jobs running on these machines. The collective performance of multiple, concurrent BFAST CU thus degrades significantly as contention for available disk I/O bandwidth increases due to larger number of tasks accessing the filesystem concurrently (due to the usage of NFS, the runtime on India is depending on the scenario $>20\%$ slower than other machines). On EGI and OSG, BFAST CU performs best. This can mainly be attributed to the use of local storage and the high degree of distribution of the CUs to multiple sites.

### C. PJ Framework Interoperability

In principal, two types of interoperability between Pilot-Jobs and infrastructure exist: the first is the usage of a given PJ framework on different infrastructures; in the scenario examined, BigJob is used on different infrastructures by invoking different SAGA adaptors. The second is the usage of distinct PJ frameworks via the Pilot-API, i.e., interoperability between PJ frameworks. In configuration C1 we utilize SAGA adaptors to run BigJob concurrently on FutureGrid:India and XSEDE:Trestles. C2 and C3 show PJ framework interoperability by concurrently running BigJob and DIANE on FutureGrid:India and EGI (C2), as well as Condor and BigJob on OSG and XSEDE:QueenBee (C3). It is worth reiterating, that to the best of our knowledge, the latter scenario, wherein different PJ frameworks are utilized concurrently for the same application (whether on the same infrastructure or distinct) has never before been realized. We attribute this to the use of the Pilot-API. For all scenarios, we run the same BFAST application described in §V-B with 64 CUs on each infrastructure, i.e. in total 128 CUs.

Figure 8 shows the results of the interoperability tests. In C1, one BJ Pilot is submitted to Trestles and one BJ Pilot to FG/India. The overall runtime is the sum of the file staging and the actual compute time on the respective resource. Consistent with previous results, the Pilot on Trestles finished before the Pilot on India. $T_R$ of the distributed run improved more than 50% compared to a run on only one resource, i.e., Trestles or India only, mainly due to the minimization of the incoming bandwidth bottleneck by distributing the load to two sites.

The middle bar (C2) in Figure 8 demonstrates that two PJ frameworks can be utilized concurrently using the Pilot-API. As previously, the overall performance heavily depends on the time required for staging the files (for configuration C2 and C3 we estimated the staging times based on previous measurements). Further, some performance overhead is induced by the distributed coordination necessary particular in case C2 where the BJ manager and Redis service are highly distributed. In this configuration, the communication is conducted via a Redis instance deployed on FG, while the BJ manager is deployed on EGI; thus, for each CU several cross-atlantic roundtrips (latencies >100 msec) are necessary. Another important aspect is file staging: as previously established, the incoming network bandwidth quickly becomes a bottleneck as in the FG case (even dominating the distributed latencies). In HTC environments the pilots and CUs, are distributed across multiple machines, which avoids such bottlenecks. Finally, C3 (right bars in Figure 8) shows the result of the OSG and XSEDE:QB run. Since QueenBee is an older XSEDE machine, $T_R$ on this machine is much longer than $T_R$ on OSG.

Although the aim of our experiments is to demonstrate interoperable use of hitherto distinct and disjoint Pilot-Jobs, in the process we highlight the performance advantages that can emanate from the ability to seamlessly distribute (I/O intensive) workloads in a scalable manner. The Pilot-API does not represent a barrier to scalability, but by virtue of facilitating the use of distributed resources, it provides the ability to overcome limitations to scalability on certain infrastructure arising from factors such as I/O, memory, and/or bandwidths.

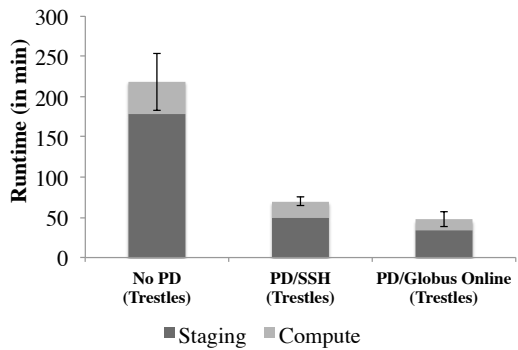

*Fig. 9:* **Pilot-Data (PD) Performance:** *Using PD the staging time can be significantly improved due to a reduction of the overall data transfer volume from 128 times 1.9 GB to about 24 GB (1.9 GB + 128 times 170 MB). Further, Globus Online utilizes with GridFTP a more efficient transfer protocol.*

### D. Pilot-Data: Data Transfer and Management

An important challenge when running distributed data-intensive tasks is the effective management of data, and its efficient transfer. We investigate different scenarios: (D1) the usage of application-level file staging without PD, (D2) the usage of PD with the SSH plugin, and (D3) the usage of PD with Globus Online. For these experiments, we utilize Trestles (XSEDE). The input files are located on a remote machine: Quarry (IU) for D1; Lonestar (TACC) for D2 and D3.

As previously alluded (Figure 8), as the number of CUs increases, file staging quickly becomes a bottleneck, as all CUs share the incoming bandwidth. Also, I/O contention on the shared file systems leads to an increasing compute time. Figure 9 shows that with an increasing number of CUs, the naive way of moving files (D1) is not a scalable solution; this is reflected in a runtime of >4 hours for 128 CUs. While the download time increases linearly with the number of CUs, the compute time remains almost constant at ∼15 min.

There are two options to address this issue: (i) distribute/scale-out the computation to multiple resources (C3), and/or (ii) optimize data transfers using PD. The optimization via PD has two components: the amount transferred and the transfer protocol. By being cognizant of the distribution of the CUs, redundant transfers can be reduced (if not eliminated); using PD the overall amount of data that needs to be transferred for the 128 CU scenario can be reduced from 243 GB (i.e. 128 CUs times 1.9 GB data for the reference genome, index files and 1 read file) to 24 GB (a single transfer of the reference genome and index files and 1 read file per CU, i.e. 128 times 170 MB). The staging times in D2 and D3 are significantly lower than in D1. Further, Globus Online show a significant better performance than SSH (∼30 %) mainly due to the usage of a more efficient transfer protocol (GridFTP).

## VI. Discussion and Future Work

The primary intellectual contributions of this work are (i) the development of the P* Model, (ii) the mapping of the P* elements to PJ frameworks such as Condor-G/Glide-in, and (iii) the design and development of the Pilot-API. The P* Model provides a common abstract model for describing and characterizing Pilot-abstractions; the Pilot-API exposes the P* elements and characteristics. We validate the P* Model by demonstrating that the most widely used PJ frameworks, viz., DIANE and Condor-G/Glide-in, can be compared, contrasted and analyzed using this analytical framework. Furthermore we demonstrate the use of the Pilot-API with multiple PJ frameworks on distributed production cyberinfrastructure, such as XSEDE, OSG, EGI and FutureGrid.

Although current PJ frameworks collectively support millions of tasks yearly on several production distributed infrastructure, extensibility and interoperability remain a significant challenge [35]. In addition, there is confusion about what constitutes a Pilot-Job system; in the absence of a well-defined model, often different semantic capabilities and functionality are compared. For example, GlideinWMS is often called a Pilot-Job, which is then logically compared to DIANE or BigJob. Using the P* Model, one can address this ambiguity and clearly establish that GlideinWMS provides a specific *scheduling* characteristic for an implementation of the Condor/Glide-in Pilot-Job.

This points to the first non-trivial deduction that this paper makes: it presents, arguably for the first time, an analytical framework upon which to construct *tools* and thereby define them as implementations of a specific and semantically well-defined capability, rather than a loosely-defined capability validated by a weak existence principle, i.e., "because it exists, it must be correct". One can argue that similar semantic tightness is required in the implementation and definition of middleware capabilities. The second non-trivial deduction is that this paper presents a conceptual framework which unifies job and data abstractions, via an extended and generalized Pilot abstraction. Given the increasing importance of Pilot-Jobs in supporting scalable dynamical execution and the challenges associated with distributed data placement, this has immense practical implications and potential.

However, the other practical implications of this work are already evident: the Pilot-API [28] has been deployed to support production-scale science on production infrastructure as emphasized in our experimentation. In fact, it is a stated goal of our research to enhance the range of applications and usage-modes that will benefit from the Pilot abstraction, by deeply integrating Pilot-API/P* capabilities with multiple production infrastructures (both grids and clouds). However, attention to several deployment issues is required, e. g., in spite of a common Pilot-API, each PJ framework has a rather different usage modality; this is reflective of the fact that typically, PJ frameworks "evolve in" and are "native to" specific infrastructure, e.g., Condor-G/Glide-in is the native PJ framework of the OSG, and its use is heavily coupled with infrastructure specific to OSG, such as GlideinWMS. This is not a limitation of our approach, but a reiteration of the need for the P* approach as a first-step in addressing deployment barriers towards interoperability.

The Pilot-Jobs concept is not limited to traditional distributed CI but also has applicability to Clouds. For example, PaaS cloud systems, such as Venus-C (Azure), support the notion of Generic Workers (worker role) which are conceptually similar to pilots in that they pull tasks (application workload) from a *central repository* when the environment is available. Furthermore, Pilot-Jobs map well to IaaS cloud systems, wherein a Pilot can marshall multiple VMs, possibly of different characteristics and performance capabilities; an agent which pulls and executes CUs is deployed on each VM. Ultimately, there is a decoupling between task specification and resource assignment, with the Pilot-Manager or an equivalent cloud-entity carrying out the mapping using dynamic/real-time information. The specifics of matching CUs to clouds is distinct from the matching CUs to grids.

P* provides significant future development & research opportunities, e. g., to experiment and reason on the relative roles of system versus application-level scheduling, heuristics for dynamic execution, the role of affinity and data/compute placement strategies, to name just a few. We will explore these ideas in upcoming work and integrate the results with production-grade implementations.

ACKNOWLEDGEMENTS

This work is funded by NSF CHE-1125332 (Cyber-enabled Discovery and Innovation), HPCOPS nsf-oci 0710874 award, NSF-ExTENCI (OCI-1007115) and NIH Grant Number P20RR016456 from the NIH National Center For Research Resources. Important funding for SAGA has been provided by the UK EPSRC grant number GR/D0766171/1 (via OMII-UK) and the Cybertools project (PI Jha) NSF/LEQSF (2007-10)-CyberRII-01, NSF EPSCoR Cooperative Agreement No. EPS-1003897 with additional support from the Louisiana Board of Regents. SJ acknowledges the e-Science Institute, Edinburgh for supporting the research theme, "Distributed Programming Abstractions" & 3DPAS. MS is sponsored by the program of BiG Grid, the Dutch e-Science Grid, which is financially supported by the Netherlands Organisation for Scientific Research, NWO. SJ acknowledges useful related discussions with Jon Weissman (Minnesota) and Dan Katz (Chicago). We thank J Kim (CCT) for assistance with BFAST. This work has also been made possible thanks to computer resources provided by TeraGrid TRAC award TG-MCB090174 (Jha) and BiG Grid. This document was developed with support from the US NSF under Grant No. 0910812 to Indiana University for "FutureGrid: An Experimental, High-Performance Grid Test-bed".